\documentclass[prb,superscriptaddress,twocolumn,showpacs,amsmath,amssymb]{revtex4}
\usepackage{graphicx} % Include figure files
\usepackage{bm}
\usepackage{txfonts}
\usepackage{mathrsfs}
\def\bp{{\mathbf p}}
\def\bq{{\mathbf q}}
\def\br{{\mathbf r}}

\def\b0{{\mathbf 0}}

\def\b0{{\mathbf 0}}

\def\cD{{\cal D}}

\def\cR{{\cal R}}
\def\cS{{\cal S}}

\def\cZ{{\cal Z}}

\def\Re{{\rm Re}}
\def\Im{{\rm Im}}
\def\bra{\langle}
\def\ket{\rangle}

\def\alf{\alpha}

\def\Gam{\Gamma}

\def\Lam{\Lambda}

\def\sg{\sigma}
\def\Sg{\Sigma}

\begin{document}

\title{Longitudinal fluctuations in the Berezinskii-Kosterlitz-Thouless phase}
\author{Pawel Jakubczyk}
\affiliation{Institute of Theoretical Physics, Faculty of Physics, University of Warsaw, Pasteura 5, 02-093 Warsaw, Poland}
\affiliation{Max-Planck-Institute for Solid State Research,
 Heisenbergstr.\ 1, D-70569 Stuttgart, Germany}
\author{Walter Metzner}
\affiliation{Max-Planck-Institute for Solid State Research,
 Heisenbergstr.\ 1, D-70569 Stuttgart, Germany} 
\date{\today}
\begin{abstract}
We analyze the interplay of longitudinal and transverse fluctuations in a $U(1)$ symmetric two-dimensional $\phi^4$-theory.
To this end, we derive coupled renormalization group equations for both types of fluctuations obtained from a linear (cartesian) decomposition of the order parameter field.
Discarding the longitudinal fluctuations, the expected Berezinskii-Kosterlitz-Thouless (BKT) phase characterized by a finite stiffness and an algebraic decay of order parameter correlations is recovered.
Renormalized by transverse fluctuations, the longitudinal mass scales to zero, so that longitudinal fluctuations become increasingly important for small momenta.
Within our expansion of the effective action, they generate a logarithmic decrease of the stiffness, in agreement with previous functional renormalization group calculations.
The logarithmic terms imply a deviation from the vanishing beta-function for the stiffness in the non-linear sigma model describing the phase fluctuations at three-loop order. 
To gain further insight, we also compute the flow of the parameters characterizing longitudinal and transverse fluctuations from a density-phase representation of the order parameter field, with a cutoff on phase fluctuations. The power-law flow of the longitudinal mass and other quantities is thereby confirmed, but the stiffness remains finite in this approach.
We conclude that the marginal flow of the stiffness obtained in the cartesian representation is an artifact of the truncated expansion of momentum dependences.
\end{abstract}


\maketitle

%%%%%%%%%%%%%%%%%%%%%%%%%%%%%%%%%%%%%%%%%%%%%%%%%%%%%%%%%%%%%%%%%%%%%%%%%%

\section{Introduction}

The Mermin-Wagner theorem \cite{mermin66} excludes the spontaneous breaking of continuous symmetries at finite temperatures in systems with reduced dimensionality $d \leq 2$.
A special situation arises for two-dimensional systems with abelian symmetry groups such as $U(1)$. In this case phase fluctuations indeed prevent long-range order at $T>0$, but the order-parameter correlations decay only algebraically at low temperatures. A phase transition to the high temperature phase with exponentially decaying correlations is driven by topological excitations, that is, vortices, as discovered independently by Berezinskii,\cite{berezinskii71} and Kosterlitz and Thouless.\cite{kosterlitz73}

There are two familiar ways of parametrizing the complex order parameter field $\phi(\br)$ in a $U(1)$-symmetric theory.
One possibility is a linear (cartesian) decomposition of the form $\phi(\br) = \alf + \sg(\br) + i\pi(\br)$, where $\alf$ is the expectation value of $\phi(\br)$, while $\sg(\br)$ and $\pi(\br)$ are real fields, usually refered to as {\em longitudinal} and {\em transverse} components. The massless transverse fluctuations lead to infrared divergences in perturbation theory, \cite{popov87} which require a renormalization group treatment.
Alternatively one may decompose the complex field as $\phi(\br) = A(\br) \, e^{i\vartheta(\br)}$ in polar coordinates, with an {\em amplitude} $A(\br)$ and a {\em phase} $\vartheta(\br)$. Writing the amplitude as a square root of {\em density}, $A(\br) = \sqrt{\rho(\br)}$, the $\phi^4$-interaction is transformed to a quadratic density-fluctuation term, while interaction terms are generated from the quadratic $|\nabla\phi|^2$ term. The latter are however suppressed at small momenta, so that no infrared divergences arise in this representation. \cite{popov87}

Usually the BKT-transition is described by phase fluctuations.
If amplitude fluctuations \cite{higgs} are present, such as in an interacting Bose gas, they are expected to be innocuous due to their finite mass.
The coupled system of longitudinal and transverse fluctuations has been analyzed in a series of studies \cite{grater95,gersdorff01,jakubczyk14} via the functional renormalization group (fRG), also known as exact or non-perturbative renormalization group. \cite{berges02,delamotte12,metzner12}
This approach is based on an exact flow equation for an effective action $\Gam^{\Lam}[\phi]$, which, as a function of an infrared cutoff $\Lam$, interpolates continuously between the bare action $\cS[\phi]$ and the generating functional for one-particle irreducible vertex functions $\Gam[\phi]$.\cite{wetterich93}
Using a non-perturbative derivative expansion,\cite{berges02,delamotte12,canet03} signatures of the BKT-transition were obtained,\cite{gersdorff01,jakubczyk14} although vortices do not appear explicitly in that approach.
The essential singularity of the correlation length $\xi \sim e^{c/\sqrt{T-T_{\rm BKT}}}$ above the critical temperature was reproduced, except in the immediate vicinity of $T_{\rm BKT}$.
In the low temperature phase below $T_{\rm BKT}$ the expected algebraic decay of order parameter correlations was obtained over many length scales, but at very large distances the correlations ultimately decay exponentially, with a huge but finite correlation length.\cite{gersdorff01} In the fRG flow this behavior is reflected by a quasi-fixed point of the scale-dependent phase stiffness $J^{\Lam}$ at intermediate scales $\Lam$, while ultimately $J^{\Lam}$ scales to zero for $\Lam \to 0$.
In a recent reexamination of the fRG flow equations it was found that the expected essential scaling of $\xi$ could be reproduced arbitrarily close to $T_{\rm BKT}$ by {\em fine-tuning}\/ the infrared cutoff function.\cite{jakubczyk14}
By the same procedure, the algebraic decay below $T_{\rm BKT}$ could be maintained at arbitrarily long distances, but only as long as $T$ is not too low.
The fRG has also been shown to be a powerful tool for accurate calculations of non-universal properties such as the specific heat, which exhibits a pronounced maximum above the BKT transition temperature. \cite{jakubczyk16}

The failure in obtaining a stable fixed point for the stiffness at low temperatures has been attributed to the approximate truncation of the effective action.
\cite{grater95,gersdorff01,jakubczyk14}
It is striking that a rather sophisticated approximation, which captures the subtle and peculiar features of the BKT-transition over a wide range of scales, fails to reproduce the fixed point at low temperatures, which can be easily obtained from a Gaussian theory of phase fluctuations.\cite{chaikin95}

In this article we clarify the mechanism that spoils the fixed point of the stiffness in the fRG flow. We focus on the low temperature regime well below the BKT-temperature, where vortices play no role.
Using a truncation of the effective action at quartic order in the field, with a cartesian decomposition in longitudinal and transverse components, we obtain an analytic understanding of the behavior of the flow.
The expected fixed point with an algebraic decay of order parameter correlations is recovered if longitudinal fluctuations are discarded.
However, transverse fluctuations lead to a decreasing longitudinal mass and, as an inevitable consequence, to a slowly decreasing stiffness, if longitudinal fluctuations are taken into account. Although the flow of the stiffness is only logarithmic, it ultimately results in a finite correlation length at any temperature $T>0$. This is clearly at odds with the absence of any singular corrections to the phase stiffness in the density-phase representation.
Therefore, we relate the longitudinal-transverse and the density-phase representation to each other by computing the flow of parameters in the cartesian decomposition from a density-phase representation with a cutoff on phase fluctuations. Thereby, the flow of the longitudinal mass to zero is confirmed, while the stiffness remains finite.

The paper is structured as follows.
In Sec.~II we introduce the $U(1)$-symmetric $\phi^4$-model and describe the ansatz for the truncated effective action.
The corresponding flow equations are presented in Sec.~III.
In Sec.~IV we analyze the flow of the stiffness and other quantities, first without, and then with longitudinal fluctuations.
In Sec.~V we revisit the flows from a density-phase representation perspective.
In the final Sec.~VI we summarize and conclude that the instability of the BKT fixed point observed in the cartesian representation is most likely an artifact of the truncation.

%%%%%%%%%%%%%%%%%%%%%%%%%%%%%%%%%%%%%%%%%%%%%%%%%%%%%%%%%%%%%%%%%%%%%%%%%%

\section{Bare and effective action}

We consider a two-dimensional system with a complex order parameter field $\phi(\br)$ described by a $U(1)$-symmetric action of the form
\begin{equation} \label{cS}
 \cS[\phi] = 
 \beta \, \left[ \frac{u_0}{8} \int d^2\br \, 
 \left( |\phi(\br)|^2 - \alf_0^2 \right)^2
 + \frac{Z_0}{2} \int d^2\br \, |\nabla\phi(\br)|^2 \right] \; ,
\end{equation}
where $u_0$, $\alf_0$, and $Z_0$ are real positive constants, and $\beta = 1/T$ is the inverse temperature.
The potential in the first term has the form of a Mexican hat with a degenerate minimum at any complex $\phi$ with modulus $\alf_0$.
Hence, the bare action exhibits spontaneous symmetry breaking.
The system is regularized by an ultraviolet momentum cutoff $\Lam_{\rm uv}$.
Note that we analyze only thermal, not quantum fluctuations.

A scale-dependent effective action \cite{wetterich93} is defined by adding an infrared regulator
\begin{equation}
 \cR^\Lam[\phi] = \frac{T}{2} \int_\bq R^\Lam(\bq) \, \phi_\bq^* \phi_\bq \; ,
\end{equation}
with a suitable regulator function $R^\Lam(\bq)$ to the bare action.
Here and in the following $\phi_\bq$ is the order parameter field in momentum representation, with normalization conventions such that
$\phi(\br) = T \int_\bq \phi_\bq \, e^{i\bq \cdot \br}$,
and $\int_\bq$ is a short-hand notation for $\int \frac{d^2\bq}{(2\pi)^2}$.
The partition function with complex source fields $h$ is then given by the functional integral
\begin{equation}
 \cZ^\Lam[h] = \int \cD[\phi] e^{-\cS[\phi] - \cR^\Lam[\phi] +
 \int_\bq (h_\bq \phi_\bq^* + h_\bq^* \phi_\bq)} \, .
\end{equation}
The Legendre transform of $-\ln\cZ^\Lam[h]$ yields the scale-dependent effective action
\begin{equation}
 \Gam^\Lam[\phi] = - \ln \cZ^\Lam[h] +
 \int_\bq (h_\bq \phi_\bq^* + h_\bq^* \phi_\bq) - \cR^\Lam[\phi] \, .
\end{equation}
The regulator $\cR^\Lam[\phi]$ suppresses fluctuations with $|\bq| < \Lam$.
For $\Lam = \Lam_0$, all fluctuations are completely suppressed such that $\Gam^{\Lam_0}[\phi] = \cS[\phi]$, which defines the initial condition of the flow.
For a sharp momentum cutoff one can choose $\Lam_0 = \Lam_{\rm uv}$, while for a smooth regulator function $R^{\Lam}(\bq)$ the initial flow parameter is $\Lam_0 = \infty$.
For $\Lam \to 0$, the regulator function $R^\Lam(\bq)$ vanishes and $\Gam^\Lam[\phi]$ tends to the final effective action $\Gam[\phi]$.
The flow of $\Gam^\Lam[\phi]$ is determined by the exact equation \cite{wetterich93}
\begin{equation} \label{exactflow}
 \partial_\Lam \Gam^\Lam[\phi] = \frac{1}{2}
 {\rm tr} \frac{\partial_\Lam R^{\Lam}}{\Gam^{(2)\Lam}[\phi] + R^\Lam} \, ,
\end{equation}
where $\Gam^{(2)\Lam}[\phi]$ is the matrix of second functional derivatives of $\Gam^{\Lam}[\phi]$ with respect to $\phi$ and $\phi^*$.

The exact effective action is a complicated functional of $\phi$.
In Refs.\ \onlinecite{gersdorff01} and \onlinecite{jakubczyk14}, it was approximated by a derivative expansion of the form (with slightly different notations)
\begin{eqnarray} \label{Gamderiv}
 \Gam^\Lam[\phi] &=& 
 \beta \int d^2\br \, \Big\{ U^\Lam[\rho(\br)] +
 \frac{1}{2} Z^\Lam[\rho(\br)] \, |\nabla\phi(\br)|^2 \nonumber \\
 && + \, \frac{1}{8} Y^\Lam[\rho(\br)] \left( \nabla |\phi(\br)|^2 \right)^2
 \Big\} \; ,
\end{eqnarray}
where $U^\Lam$, $Z^\Lam$, and $Y^\Lam$ are functions of the $U(1)$-invariant
$\rho(\br) = |\phi(\br)|^2$.
The second gradient term is important for capturing the distinct longitudinal and transverse gradients.\cite{tetradis94}
The unrestricted $\rho$-dependence of $U^\Lam$, $Z^\Lam$, and $Y^\Lam$ turned out to be a crucial ingredient in the BKT-transition regime.\cite{gersdorff01,jakubczyk14}
At low temperatures, however, one may discard the $\rho$ dependence of $Z^\Lam$ and $Y^\Lam$, and truncate $U^\Lam$ at quadratic order in $\rho$, that is, at quartic order in $\phi$.
We are thus led to our ansatz for the effective action,
\begin{eqnarray} \label{Gam}
 \Gam^\Lam[\phi] &=& 
 \beta \int d^2\br \, \left[ \frac{u^\Lam}{8} \,
 \left( |\phi(\br)|^2 - (\alf^\Lam)^2 \right)^2 \right. \nonumber \\ 
 &+& \left. \frac{Z^\Lam}{2} \, |\nabla\phi(\br)|^2 + 
 \frac{Y^\Lam}{8} \,
 \left( \nabla |\phi(\br)|^2 \right)^2 \right] \; ,
\end{eqnarray}
with positive scale-dependent numbers $u^\Lam$, $\alf^\Lam$, $Z^\Lam$, and $Y^\Lam$.
The first two terms have the form of the bare action, with renormalized parameters.
In the following we will drop the $\Lam$-superscripts from scale-dependent quantities, to simplify the notation.

To expand around the real positive minimum of the mexican hat potential, we decompose $\phi(\br)$ linearly (in cartesian coordinates) as
\begin{equation} \label{cartesian}
 \phi(\br) = \alf + \sg(\br) + i\pi(\br) \; ,
\end{equation}
where $\sg(\br)$ and $\pi(\br)$ are real fields describing longitudinal and transverse fluctuations, respectively. Their Fourier components $\sg_\bq$ and $\pi_\bq$ obey the relations $\sg_\bq^* = \sg_{-\bq}$ and $\pi_\bq^* = \pi_{-\bq}$.
Inserting the decomposition of $\phi$ into Eq.~(\ref{Gam}),
one obtains several quadratic, cubic and quartic terms,
\begin{equation}
 \Gam =
 \Gam_{\sg^2} + \Gam_{\pi^2} + \Gam_{\sg^3} + \Gam_{\sg\pi^2} +
 \Gam_{\sg^4} + \Gam_{\pi^4} + \Gam_{\sg^2\pi^2} \; .
\end{equation}
The quadratic terms have the form
\begin{eqnarray}
 \Gam_{\sg^2} &=& \frac{T}{2} \int_{\bq}
 \big( m_{\sg}^2 + Z_{\sg} \, \bq^2 \big) \, \sg_\bq \sg_{-\bq} \; ,
 \nonumber \\
 \Gam_{\pi^2} &=& \frac{T}{2} \int_\bq
 Z_{\pi} \, \bq^2 \, \pi_\bq \pi_{-\bq} \; .
\end{eqnarray}
The longitudinal mass is determined by the order parameter $\alf$ and the quartic coupling $u$ in Eq.~(\ref{Gam}) as
\begin{equation} \label{rel_mualf}
 m_{\sg}^2 = u \alf^2 \; ,
\end{equation}
and the $Z$-factors for longitudinal and transverse fluctuations are related to $Z$ and $Y$ by
\begin{eqnarray} \label{rel_ZY}
 Z_{\sg} &=& Z + Y \alf^2 \; , 
 \nonumber \\
 Z_{\pi} &=& Z \; .
\end{eqnarray}
We now see that the quartic gradient term in the action (\ref{Gam}) is crucial for allowing independent renormalizations of longitudinal and transverse fluctuations.\cite{tetradis94}
The cubic and quartic interaction terms read
\begin{eqnarray} \label{rel_vertex}
 \Gam_{\sg^3} &=& \frac{T^2}{2} \int_{\bq,\bp} 
 U(\bp) \, \alf \, \sg_\bp \sg_\bq \sg_{-\bq-\bp} \; , \nonumber \\
  \Gam_{\sg\pi^2} &=& \frac{T^2}{2} \int_{\bq,\bp} 
 U(\bp) \, \alf \, \sg_\bp \pi_\bq \pi_{-\bq-\bp} \; , \nonumber \\
 \Gam_{\sg^4} &=& \frac{T^3}{8} \int_{\bq,\bq',\bp}
 U(\bp) \, \sg_\bq \sg_{\bp-\bq} \sg_{\bq'} \sg_{-\bp-\bq'} \; , \nonumber \\
 \Gam_{\pi^4} &=& \frac{T^3}{8} \int_{\bq,\bq',\bp}
 U(\bp) \, \pi_\bq \pi_{\bp-\bq} \pi_{\bq'} \pi_{-\bp-\bq'} \; , \nonumber \\
 \Gam_{\sg^2\pi^2} &=& \frac{T^3}{4} \int_{\bq,\bq',\bp}
 U(\bp) \, \sg_\bq \sg_{\bp-\bq} \pi_{\bq'} \pi_{-\bp-\bq'} \; ,
\end{eqnarray}
where $U(\bp) = u + Y \bp^2$.
The relations (\ref{rel_mualf})-(\ref{rel_vertex}) are valid for the quartic action (\ref{Gam}). Higher order terms would yield additional contributions.

A quantum version of the ansatz (\ref{Gam}) for the effective action fully captures the subtle interplay of longitudinal and transverse fluctuations in the ground state of an interacting Bose gas.\cite{obert13}

%%%%%%%%%%%%%%%%%%%%%%%%%%%%%%%%%%%%%%%%%%%%%%%%%%%%%%%%%%%%%%%%%%%%

\section{Flow equations}

Inserting the ansatz (\ref{Gam}) for the effective action $\Gam^\Lam$ into the exact flow equation (\ref{exactflow}) and comparing coefficients, one obtains flow equations for the scale-dependent parameters.
The parameters $u$, $\alf$, $Z$, and $Y$ in Eq.~(\ref{Gam}) are related to the longitudinal mass $m_\sg^2$ and the $Z$-factors for longitudinal and transverse fluctuations, $Z_\sg$ and $Z_\pi$, by the relations (\ref{rel_mualf}) and (\ref{rel_ZY}).
We can thus avoid flow equations for four-point functions, and use $\alf$, $m_\sg^2$, $Z_\sg$, and $Z_\pi$ as our basic variables. The flow of $m_\sg^2$, $Z_\sg$, and $Z_\pi$ is determined from the self-energies for $\sg$- and $\pi$-fields, that is, two-point functions.

The right-hand sides of the flow equations are loop integrals with interaction vertices and $\sg$- and $\pi$-propagators
\begin{eqnarray}
\label{G_sg}
 G_\sg(\bq) &=& \bra \sg_\bq \sg_{-\bq} \ket =
 \frac{1}{m_\sg^2 + Z_\sg \bq^2 + R(\bq)} \, , \\
\label{G_pi}
 G_\pi(\bq) &=& \bra \pi_\bq \pi_{-\bq} \ket =
 \frac{1}{Z_\pi \bq^2 + R(\bq)} \, .
\end{eqnarray}
The derivation of the flow-equations is straightforward. In the present case, one can obtain them also by expanding the flow equations for the derivative expansion used in Ref.~\onlinecite{jakubczyk14}, or, even more easily, by extracting the classical (zero frequency) bosonic contributions from the flow equations in Ref.~\onlinecite{obert13}.

The flow equation for $\alf$ is obtained from the condition that the one-point $\sg$-vertex has to vanish if $\alf$ is the modulus of the minimum of the effective action.
This yields \cite{obert13}
\begin{equation} \label{floweq_alf}
 \frac{d\alf}{d\Lam} = - \frac{\alf T}{2m_\sg^2} \int_\bq 
 \big\{ [u + 2U(\bq)] G'_\sg(\bq) + u G'_\pi(\bq) \big\} \, .
\end{equation}
Here and in the following the prime indicates a $\Lam$-derivative acting only on $R(\bq)$, that is,
\begin{equation} \label{DLam}
 G'_{\sg/\pi}(\bq) = D_\Lam G_{\sg/\pi}(\bq) =
 - G_{\sg/\pi}^2 \frac{dR(\bq)}{d\Lam} \, .
\end{equation}
The flow equations for $m_\sg^2$ and $Z_\sg$ can be derived from the flow equation for the self-energy of the $\sg$-fields, which has the form \cite{obert13}
\begin{eqnarray} \label{floweq_Sg_sg}
 \frac{d}{d\Lam} \Sg_{\sg}(\bp) &=& 
 \left[ u + 2U(\bp) \right] \alf \frac{d\alf}{d\Lam}
 \nonumber \\
 &+& \frac{T}{2} \int_\bq 
 \left\{ \left[ u + 2U(\bp+\bq) \right] \, G'_{\sg}(\bq) 
 + u \, G'_{\pi}(\bq) \right\}
 \nonumber \\
 &-& \frac{T}{2} \int_\bq 
 \left[ U(\bp) + U(\bq) + U(\bp+\bq) \right]^2 \alf^2
 \nonumber \\
 &\times& D_{\Lam} \left[ G_{\sg}(\bq) G_{\sg}(\bp+\bq) \right]
 \nonumber \\
 &-& \frac{T}{2} \int_\bq 
 \left[ U(\bp) \right]^2 \alf^2 \,
 D_{\Lam} \left[ G_{\pi}(\bq) G_{\pi}(\bp+\bq) \right] ,
\end{eqnarray}
with $D_\Lam$ as defined in Eq.~(\ref{DLam}).
The flow of the longitudinal mass $m_\sg^2 = m_{\sg 0}^2 + \Sg_\sg(\b0)$ is thus obtained from
\begin{eqnarray} \label{floweq_m_sg}
 \frac{d}{d\Lam} m_{\sg}^2 &=& 
 3 u \alf \frac{d\alf}{d\Lam} +
 \frac{T}{2} \int_\bq 
 \left\{ \left[ u + 2U(\bq) \right] \, G'_{\sg}(\bq) 
 + u \, G'_{\pi}(\bq) \right\} \quad
 \nonumber \\
 &-& \frac{\alf^2 T}{2} \int_\bq D_{\Lam}
 \left\{ \left[ u + 2U(\bq) \right]^2 G_{\sg}^2(\bq)
 + u^2 \, G_{\pi}^2(\bq) \right\} .
\end{eqnarray}
The flow of $Z_{\sg}$ can be obtained from a second momentum
derivative of the self-energy at $\bp=0$, that is,
$\frac{d}{d\Lam} Z_{\sg} = \left. \frac{1}{2} \partial_{p_x}^2
 \frac{d}{d\Lam} \Sg_{\sg}(\bp) \right|_{\bp=\b0} =
 \left. \frac{1}{4} \Delta_\bp
 \frac{d}{d\Lam} \Sg_{\sg}(\bp) \right|_{\bp=\b0} \,$, where
$\Delta_\bp = \partial_{p_x}^2 + \partial_{p_y}^2$, yielding
\begin{eqnarray} \label{floweq_Z_sg}
 \frac{d}{d\Lam} Z_{\sg} &=& 
 2Y \alf \frac{d\alf}{d\Lam} + T Y \int_\bq G'_{\sg}(\bq)
 \nonumber \\
 &-& \frac{T}{8} \, \Delta_\bp \int_\bq
 \left[ U(\bp) + U(\bq) + U(\bp+\bq) \right]^2 \alf^2 \,
 \nonumber \\
 &\times& D_{\Lam} \left[ G_{\sg}(\bq) G_{\sg}(\bp+\bq) \right] \,
 \big|_{\bp=0} \nonumber \\
 &-& \frac{T}{8} \Delta_\bp \int_\bq
 \left[ U(\bp) \right]^2 \alf^2
 D_{\Lam} \left[ G_{\pi}(\bq) G_{\pi}(\bp+\bq) \right]
 \big|_{\bp=0} . \quad\quad
\end{eqnarray}
The flow equation for the $\pi$-field self-energy reads \cite{obert13}
\begin{eqnarray} \label{floweq_Sg_pi}
 \frac{d}{d\Lam} \Sg_{\pi}(\bp) &=& 
 u \alf \frac{d\alf}{d\Lam} 
 \nonumber \\
 &+& \frac{T}{2} \int_\bq
 \left\{ \left[ u + 2U(\bp+\bq) \right] \, G'_{\pi}(\bq) 
 + u \, G'_{\sg}(\bq) \right\} 
 \nonumber \\
 &-& T \int_\bq 
 \left[ U(\bq) \right]^2 \alf^2 \,
 D_{\Lam} \left[ G_{\sg}(\bq) G_{\pi}(\bp+\bq) \right] \, .
\end{eqnarray}
The flow of $Z_{\pi}$ can be extracted by applying a second 
order momentum derivative, that is,
$\frac{d}{d\Lam} Z_{\pi} = \left. \frac{1}{4} \Delta_\bp 
 \frac{d}{d\Lam} \Sg_{\pi}(\bp) \right|_{\bp=\b0} \,$, yielding
\begin{eqnarray}
 \frac{dZ_\pi}{d\Lam} &=&
 \frac{T}{4} \, \Delta_\bp \int_\bq U(\bp+\bq) G'_\pi(\bq) \big|_{\bp=\b0}
 \nonumber \\
 &-& \frac{T}{4} \, \Delta_\bp \int_\bq [U(\bq)]^2 \alf^2
 D_\Lam [G_\sg(\bq) G_\pi(\bp+\bq)] \big|_{\bp=\b0} \, . \quad
\end{eqnarray}
Using the relation $\alf^2 U(\bq) = G_\sg^{-1}(\bq) - G_\pi^{-1}(\bq)$, this can be simplified to
\begin{eqnarray} \label{floweq_Z_pi}
 \frac{dZ_\pi}{d\Lam} &=&
 \frac{T}{4} \, \Delta_\bp \int_\bq U(\bq) \, D_{\Lam} \Big[
 G_{\sg}(\bq) G_\pi^{-1}(\bq) G_\pi(\bp+\bq) \Big] \Big|_{\bp=\b0}
 \nonumber \\
 &=& \frac{T}{4\alf^2} \, \Delta_\bp \int_\bq D_\Lam
 \Big[ G_\pi^{-1}(\bq) \, G_\pi(\bp+\bq)
 \nonumber \\
 &-& G_\sg(\bq) \, G_\pi^{-2}(\bq) \, G_\pi(\bp+\bq) \Big] \Big|_{\bp=0} \, .
\end{eqnarray}
In the last step the contribution from transverse fluctuations has been disentangled from contributions involving also longitudinal fluctuations.

%%%%%%%%%%%%%%%%%%%%%%%%%%%%%%%%%%%%%%%%%%%%%%%%%%%%%%%%%%%%%%%%%%%%

\section{Flow}

We now study the flow of the parameters $\alf$, $m_\sg^2$, $Z_\sg$ and $Z_\pi$ as obtained from the flow equations derived in the preceding section.
We will first show that the BKT fixed point is easily recovered if only transverse fluctuations are taken into account, and will then analyze the behavior and impact of longitudinal fluctuations.

%%%%%%%%%%%%%%%%%%%%%%%%%%%%%%%%%%%%%%%%%%%%%%%%%%%%%%%%%%%%%%%%%%%%%%%%%%%%%%

\subsection{BKT fixed point}

The flow equations involve longitudinal and transverse fluctuations.
Discarding the longitudinal fluctuations, the flow equation (\ref{floweq_alf}) for the order parameter $\alf$ is reduced to
\begin{equation}
 \frac{d\alf}{d\Lam} = - \frac{\alf T}{2m_\sg^2} \int_\bq u \, G'_\pi(\bq) \, .
\end{equation}
Using $m_\sg^2 = u\alf^2$, this can be written as
\begin{equation} \label{floweq2_alf}
 \frac{d\alf^2}{d\Lam} = - T \int_\bq G'_\pi(\bq) \, .
\end{equation}
The flow equation (\ref{floweq_Z_pi}) for $Z_\pi$ is also simplified,
\begin{equation} \label{floweq1_Z_pi}
 \frac{dZ_\pi}{d\Lam} =
 \frac{T}{4\alf^2} \, \Delta_\bp \int_\bq D_{\Lam} \left[
 G_\pi^{-1}(\bq) G_\pi(\bp+\bq) \right] \big|_{\bp=\b0} \, ,
\end{equation}
if the contribution from longitudinal fluctuations is omitted.
The simplified flow equations (\ref{floweq2_alf}) and (\ref{floweq1_Z_pi}) represent a closed system involving only $\alf$ and $Z_\pi$. The coupling $u$ and the longitudinal mass $m_\sg^2$ have canceled out.
After substituting $\bq \mapsto \bq-\bp$ in the integral in Eq.~(\ref{floweq1_Z_pi}), the momentum derivative acts on $G_\pi^{-1}$, so that one obtains
\begin{equation} \label{floweq2_Z_pi}
 \frac{dZ_\pi}{d\Lam} =
 \frac{T}{\alf^2} \, Z_\pi \int_\bq G'_\pi(\bq) +
 \frac{T}{4\alf^2} \, D_\Lam \int_\bq G_\pi(\bq) \, \Delta_\bq R(\bq) \, .
\end{equation}
For a momentum independent regulator function, such as $R = Z_\pi \Lam^2$, the last term in this flow equation vanishes.
For a generic regulator function with the usual asymptotic properties $R(\bq) \to 0$ for $\Lam \to 0$ and $R(\bq) \to {\rm const}$ for $\bq \to \b0$, the integral $\int_\bq G_\pi(\bq) \, \Delta_\bq R(\bq)$ converges to a constant for $\Lam \to 0$, such that its $\Lam$-derivative vanishes in that limit.
Hence, the second term in Eq.~(\ref{floweq2_Z_pi}) vanishes in any case for $\Lam \to 0$, such that
\begin{equation} \label{floweq3_Z_pi}
 \frac{dZ_\pi}{d\Lam} =
 \frac{T}{\alf^2} \, Z_\pi \int_\bq G'_\pi(\bq) \,
\end{equation}
in the infrared limit.

From the flow equations for $\alf$ and $Z_\pi$, Eqs.~(\ref{floweq2_alf}) and (\ref{floweq3_Z_pi}), one can immediately see that
\begin{equation}
 \frac{d}{d\Lam} (Z_{\pi} \alf^2) = 0 \, .
\end{equation}
Hence, the phase stiffness
\begin{equation}
 J = Z_\pi \alf^2
\end{equation}
tends to a finite constant for $\Lam \to 0$. The finite phase stiffness is a hallmark of the BKT phase.\cite{chaikin95} In particular, in case of an interacting Bose gas, the finite stiffness implies that the system is a superfluid at low temperatures, although there is no long-range order.

For simple regulator functions, the coupled flow equations for $\alf$ and $Z_\pi$ can be solved explicitly.
In particular, for a momentum-independent regulator $R$ one obtains 
\begin{equation}
 \int_\bq G'_\pi(\bq) = - \frac{1}{4\pi Z_\pi} \frac{d\ln R}{d\Lam}
\end{equation}
for small $\Lam$ (that is, for $R \ll Z_\pi \Lam_{\rm uv}^2$), such that 
\begin{eqnarray}
 \frac{d}{d\Lam} \alf^2 &=& \frac{T}{4\pi Z_\pi} \frac{d\ln R}{d\Lam} \, , \\
 \frac{d}{d\Lam} Z_\pi &=& - \frac{T}{4\pi \alf^2} \frac{d\ln R}{d\Lam} \, .
\end{eqnarray}
The solution of these equations has the form
$\alf^2 \propto R^{T/(4\pi J)}$ and $Z_\pi \propto R^{-T/(4\pi J)}$.
For the specific choice $R = \Lam^2$, where the parameter $\Lam$ is a momentum scale, one obtains
\begin{eqnarray}
\label{alf}
 \alf^2 &=& \bar\alf_0^2 (\Lam/\Lam_{\rm uv})^\eta \, , \\
\label{Z_pi}
 Z_\pi &=& \bar Z_{\pi0} (\Lam_{\rm uv}/\Lam)^\eta \, ,
\end{eqnarray}
with the anomalous dimension
\begin{equation} \label{eta}
 \eta = \frac{T}{2\pi J} \, .
\end{equation}
The prefactors $\bar\alf_0^2$ and $\bar Z_{\pi0}$ are renormalized counterparts of the bare parameters $\alf_0^2$ and $Z_{\pi0}$.

In previous fRG calculations the regulator $R$ was often chosen proportional to $Z_\pi$, so that common factors of $Z_\pi$ could be scaled away by a redefinition of variables.
For $R = Z_\pi \Lam^2$, the transverse propagator has the simple form $G_\pi^{-1}(\bq) = Z_\pi (\bq^2 + \Lam^2)$.
Since $R$ depends on $Z_\pi$ in this case, it appears also on the right hand side of the above solution as a function of $R$. Solving for $\alf$ and $Z_\pi$ as a function of $\Lam$, one again obtains a solution of the form Eqs.~(\ref{alf}) and (\ref{Z_pi}), where now
\begin{equation} \label{eta_mod}
 \eta = \frac{T/(2\pi J)}{1 + T/(4\pi J)} \approx \frac{T}{2\pi J} \;
 \mbox{for} \; T \ll 4\pi J \, .
\end{equation}
Note that the difference between this equation and Eq.~(\ref{eta}) is not a contradiction, but simply reflects the different choice of the flow parameter $\Lam$.

The order parameter $\alf$ vanishes for $\Lam \to 0$, in agreement with the Mermin-Wagner theorem. $Z_\pi$ diverges with a power-law for $\Lam \to 0$, which implies a modification of the quadratic momentum dependence of the transverse fluctuation propagator to
$G_\pi(\bq) \sim |\bq|^{\eta - 2}$ for small $\bq$. This corresponds to the well-known algebraic decay of order parameter correlations at long distances, with an exponent proportional to $T$ for low $T$.\cite{chaikin95}

%%%%%%%%%%%%%%%%%%%%%%%%%%%%%%%%%%%%%%%%%%%%%%%%%%%%%%%%%%%%%%%%%%%%%%%

\subsection{Longitudinal fluctuations at the BKT-fixed point}

We now analyze how the transverse fluctuations affect the mass and the $Z$-factor of the longitudinal fluctuations. In the ground state of a superfluid Bose gas, the transverse quantum fluctuations lead to substantial renormalizations of the longitudinal fluctuations in $d \leq 3$ dimensions: $m_\sg^2$ vanishes and $Z_\sg$ diverges in the low-energy limit. \cite{nepomnyashchii75, pistolesi04, dupuis09, obert13}
In this section we still assume that the transverse fluctuations yield the dominant contributions to the flow, so that the flow of $\alf$ and $Z_\pi$ is described by the BKT fixed-point as discussed above.
Although this assumption turns out to be inconsistent with its consequences, the power-laws derived below remain valid in a broad scaling regime at low temperatures.

The dominant  transverse fluctuation contribution to the flow of the mass $m_\sg^2$ is given by the last term in Eq.~(\ref{floweq_m_sg}). Discarding the other terms and using $u  = m_\sg^2/\alf^2$ one obtains
\begin{equation}
 \frac{d}{d\Lam} m_\sg^2 = - T \frac{m_\sg^4}{\alf^2} \int_\bq
 G_\pi(\bq) \, G'_\pi(\bq) \, ,
\end{equation}
which can also be written as
\begin{equation} \label{floweq2_m_sg}
 \frac{d}{d\Lam} (m_\sg^2)^{-1} = \frac{T}{\alf^2} \int_\bq
 G_\pi(\bq) \, G'_\pi(\bq) \, .
\end{equation}
Instead of solving the flow equation for a specific choice of regulator, we perform a general power-counting analysis for an arbitrary regulator that suppresses contributions from momenta below the cutoff scale $\Lam$.
At the BKT fixed point, one has
$\alf^2 \sim \Lam^\eta$, $Z_\pi \sim \Lam^{-\eta}$, and thus
$G_\pi \sim \Lam^{\eta - 2}$, $G'_\pi \sim \Lam^{\eta - 3}$.
Hence, for small $\Lam$, the flow equation Eq.~(\ref{floweq2_m_sg}) has the form $\partial_\Lam (m_\sg^2)^{-1} = - T C \Lam^{\eta - 3}$,
where $C > 0$ is a constant (note that $G'_\pi$ is negative).
This equation can be easily integrated, yielding
\begin{equation} \label{m_sg}
 m_\sg^2 = \frac{2-\eta}{C T} \, \Lam^{2-\eta}
\end{equation}
for small $\Lam$.
The initial mass $m_{\sg0}^2$ does not enter here, the constant $C$ depends only on $\alf$ and $Z_\pi$.
The longitudinal mass thus vanishes rapidly for $\Lam \to 0$, with an exponent $2-\eta$, and a prefactor proportional to $T^{-1}$.
Note that $\eta \leq 1/4$ in the BKT phase,\cite{chaikin95} so that $2-\eta$ remains close to $2$.
From $u = m_\sg^2/\alf^2$ it follows that the $\phi^4$-coupling scales as
$u \sim T^{-1} \Lam^{2-2\eta}$, and thus vanishes, too.

Keeping only the transverse fluctuation term, the flow equation (\ref{floweq_Z_sg}) for $Z_\sg$ can be simplified to
\begin{eqnarray}
 \frac{d}{d\Lam} Z_{\sg} &=& 
 - \frac{T}{8} \, \Delta_\bp \int_\bq \left.
 \left[ U(\bp) \right]^2 \alf^2 \,
 D_{\Lam} \left[ G_{\pi}(\bq) G_{\pi}(\bp+\bq) \right] \,
 \right|_{\bp=0} \nonumber \\
 &=& -T u (Z_\sg - Z_\pi) \, D_\Lam \int_\bq G_\pi^2(\bq)
 \nonumber \\
 &-& \frac{T}{8} u^2 \alf^2 D_\Lam \int_\bq G_\pi(\bq) \, \Delta_\bq G_\pi(\bq) \, .
\end{eqnarray}
In the last step we have used the relation (\ref{rel_ZY}) to replace $Y\alf^2$ by $Z_\sg - Z_\pi$. At the BKT fixed point, one has
$u \sim \Lam^{2-2\eta}$, $\alf^2 \sim \Lam^\eta$, and $G_\pi \sim \Lam^{\eta-2}$,
with $T$-independent prefactors, and $u \sim  T^{-1} \Lam^{2-2\eta}$.
Hence, for small $\Lam$, the flow equation has the form
$\partial_\Lam Z_\sg = K_1 (Z_\sg - Z_\pi) \Lam^{-1} - K_2 T^{-1} \Lam^{-1-\eta}$
with $Z_\pi = C_\pi \Lam^{-\eta}$, where $K_1$, $K_2$ and $C_\pi$ are positive $T$-independent constants.
The solution for $Z_\sg$ has the form $Z_\sg = C_\sg \Lam^{-\eta}$, with the prefactor $C_\sg$ determined by $-\eta C_\sg = K_1 (C_\sg - C_\pi) - K_2/T$, so that
\begin{equation} \label{Z_sg}
 Z_\sg = \frac{K_1 C_\pi + K_2/T}{K_1 + \eta} \Lam^{-\eta}
 \to \frac{K_2}{K_1} T^{-1} \Lam^{-\eta}
\end{equation}
for low temperatures.
Hence, $Z_\sg$ diverges with the same power in $\Lam$ as $Z_\pi$, albeit with a larger prefactor proportional to $T^{-1}$ for low temperatures.
The relation $Y = (Z_\sg - Z_\pi)/\alf^2$ then implies that the quartic gradient coupling $Y$ diverges as $Y \sim T^{-1} \Lam^{-2\eta}$.

Let us now check the assumptions underlying the derivation of the BKT fixed point in Sec.~IV.A. We assumed that transverse fluctuations dominate over longitudinal fluctuations for small $\Lam$, so that the contributions from terms involving $G_\sg$ could be discarded. Indeed $G_\sg(\bq)$ is initially suppressed by the longitudinal mass $m_\sg^2$.
However, we have just shown that this mass vanishes as $\Lam^{2-\eta}$ at the BKT fixed point. Furthermore, $Z_\sg \sim \Lam^{-\eta}$, such that $Z_\sg \bq^2$ also vanishes as $\Lam^{2-\eta}$.
Hence, the longitudinal and transverse propagators $G_\sg$ and $G_\pi$ actually exhibit the same power-law scaling $G_{\sg/\pi} \sim \Lam^{\eta-2}$ for small $\Lam$, so that the longitudinal flucuations are {\em not}\/ neglibible compared to the transverse fluctuations, contrary to what we assumed.

%%%%%%%%%%%%%%%%%%%%%%%%%%%%%%%%%%%%%%%%%%%%%%%%%%%%%%%%%%%%%%%%%%%%%%%

\subsection{Coupled longitudinal and transverse fluctuations}

Since the assumptions on the irrelevance of longitudinal fluctuations at the BKT fixed point turned out to be violated, we have to deal with the full coupled system of longitudinal and transverse fluctuations.
We will first show analytically that in the presence of longitudinal fluctuations the renormalized stiffness decreases logarithmically as a function of $\Lam$, that is, the BKT fixed point is marginally unstable at any temperature within the ansatz (\ref{Gam}). We then present a numerical solution of the flow equations for specific parameters. At low temperatures the BKT phase is spoiled only in the extreme infrared limit, corresponding to a finite but huge correlation length.

%%%%%%%%%%%%%%%%%%%%%%%%%%%%%%%%%%%%%%%%%%%%%%%%%%%%%%%%%%%%%%%%%%%%%%%%%%%%%%%%%%

\subsubsection{Flow of stiffness}

Under the assumptions made in Sec.~IV.A, the stiffness $J = Z_\pi \alf^2$ converges to a finite constant for $\Lam \to 0$. We now show that the coupled system of longitudinal and transverse fluctuations actually generates a logarithmic decrease of $J$.

Inserting $m_\sg^2 = u \alf^2$, the flow equation (\ref{floweq_alf}) for $\alf$ can be written as
\begin{equation}
 \frac{d\alf^2}{d\Lam} =
 - T \int_\bq \left[ G'_\pi(\bq) + G'_\sg(\bq) \right] -
 \frac{2T}{u} \int_\bq U(\bq) \, G'_\sg(\bq) \, .
\end{equation}
The first term in the flow equation (\ref{floweq_Z_pi}) for $Z_\pi$ has the same form as the right hand side of Eq.~(\ref{floweq1_Z_pi}) in Sec.~IV.A, and thus tends to the right hand side of Eq.~(\ref{floweq3_Z_pi}) for small $\Lam$. In the second term we pull the momentum-derivative under the integral. This yields
\begin{equation}
 \frac{dZ_\pi}{d\Lam} = \frac{T}{\alf^2} \, Z_\pi \int_\bq G'_\pi(\bq) -
 \frac{T}{4\alf^2} \int_\bq D_\Lam \left[ G_\sg(\bq) \, G_\pi^{-2}(\bq) \,
 \Delta_\bq G_\pi(\bq) \right] \, .
\end{equation}

In the flow equation for the stiffness only the pure transverse fluctuation terms cancel, such that
\begin{eqnarray} \label{floweq_J}
 \frac{dJ}{d\Lam} &=&
 - T Z_\pi \int_\bq \left[ 1 + 2u^{-1}U(\bq) \right] G'_\sg(\bq) \nonumber \\
 &-& \frac{T}{4} \int_\bq D_\Lam \left[ G_\sg(\bq) \, G_\pi^{-2}(\bq) \,
 \Delta_\bq G_\pi(\bq) \right] \, .
\end{eqnarray}
In the regime governed by BKT-scaling, the propagators scale as $G_\sg \sim \Lam^{\eta-2}$ and $G_\pi \sim \Lam^{\eta-2}$. Hence, all the terms on the right hand side of Eq.~(\ref{floweq_J}) are proportional to $\Lam^{-1}$ for small $\Lam$. The anomalous dimension $\eta$ cancels.
The right hand side of the flow equation (\ref{floweq_J}) is positive. Let us show this for the case of a momentum independent regulator $R$. The momentum derivative acting on $G_\pi(\bq)$ can then be carried out explicitly, yielding
$G_\pi^{-2}(\bq) \, \Delta_\bq G_\pi(\bq) = -4Z_\pi + 8 Z_\pi^2 \bq^2 G_\pi(\bq)$.
Eq.~(\ref{floweq_J}) can then be simplified to
\begin{equation} \label{floweq2_J}
 \frac{dJ}{d\Lam} =
 - 2T Z_\pi u^{-1} \! \int_\bq U(\bq) \, G'_\sg(\bq) -
 2T Z_\pi^2 D_\Lam \! \int_\bq \bq^2 G_\sg(\bq) \, G_\pi(\bq) .
\end{equation}
Now it is obvious that the right hand side of the flow equation is positive, since $G_\sg$ and $G_\pi$ decrease with increasing $\Lam$ (i.e., increasing $R$).

Hence, the scale-derivative of the stiffness is proportional to $\Lam^{-1}$, with a positive prefactor $A(T)$. The temperature dependence of the prefactor can be read off from Eq.~(\ref{floweq_J}). In addition to the explicit factor $T$ in front of the integrals, there is also another $T$-dependence entering via the propagators. While $G_\sg$ and $G_\pi$ both scale as $\Lam^{\eta-2}$, the longitudinal propagator $G_\sg$ is suppressed due to an additional enhancement of $Z_\sg$ by a factor $T^{-1}$, see Eq.~(\ref{Z_sg}). For low temperatures, one thus finds $A(T) \propto T Z_\pi/Z_\sg \propto T^2$.
Hence, for low temperatures, the flow equation for the stiffness has the form
\begin{equation} \label{floweq3_J}
 \frac{dJ}{d\Lam} = \frac{a T^2}{\Lam} \, ,
\end{equation}
with a positive constant $a > 0$. This yields a logarithmic decrease of the stiffness
\begin{equation} \label{J_log}
 J = J_0 - a T^2 \ln(\Lam_{\rm uv}/\Lam) \, .
\end{equation}
The scale at which the stiffness vanishes can thus be estimated as
\begin{equation} \label{lambda_c}
 \Lam_c = \Lam_{\rm uv} e^{-J_0/(aT^2)} \, .
\end{equation}
It vanishes exponentially for $T \to 0$. The inverse critical momentum scale can be identified with a correlation length, that is, $\xi \sim \Lam_c^{-1}$. Hence, the correlation length is finite at any temperature $T>0$, but huge at low temperatures.

The slow decrease of the stiffness as a function of the scale in a $U(1)$-symmetric $\phi^4$-theory in two dimensions was first observed by Gr\"ater and Wetterich. \cite{grater95} They applied a quartic truncation of the effective action similar to ours, but without the quartic gradient term, so that the $Z$-factors for longitudinal and transverse fluctuations remained identical in their truncation.
In this approximation $G_\sg$ is not suppressed by an additional factor $T$ compared to $G_\pi$, so that the flow of the stiffness $dJ/d\Lam$ is proportional to $T$ instead of $T^2$. Hence, improving the truncation by adding the quartic gradient term reduces the flow of the stiffness.

It is instructive to compare the above results to the loop expansion of the non-linear sigma model. \cite{amit84} The non-linear sigma model describes only phase fluctuations. It requires only two renormalization parameters, the coupling constant $g = K^{-1}$ with $K = J/T$, and the wave function renormalization $Z$. In two dimensions, the loop expansion of the beta-function $\beta_g$ determining the flow of $g$ corresponds to an expansion in powers of $g$. For the abelian symmetry group $U(1)$, all contributions to $\beta_g$ vanish. Gr\"ater and Wetterich \cite{grater95} pointed out that their truncation is consistent with $\beta_g$ to one-loop order (order $g^2$), but a discrepancy arises at the two-loop level. A similar behavior was obtained for a relativistic quantum $O(N)$ model, but the correct one-loop result was reproduced  only for a suitable cutoff function. \cite{rancon13} In our improved truncation the inconsistency is shifted by one order to the three-loop level, independently of the cutoff choice.

Although the BKT fixed point turned out to be marginally unstable, at low temperatures the power-laws derived in Secs.~IV.A and IV.B are nevertheless valid in a wide regime, with a scale-dependent anomalous exponent $\eta$ increasing only logarithmically upon lowering $\Lam$.
This ''quasi-fixed point'' regime is seen very clearly in the numerical solution of the flow equations presented in the following section.

%%%%%%%%%%%%%%%%%%%%%%%%%%%%%%%%%%%%%%%%%%%%%%%%%%%%%%%%%%%%%%%%%%%%%%%%%%%%%%%%%

%
\begin{figure}[tb]
\begin{center}
\includegraphics[width=7cm]{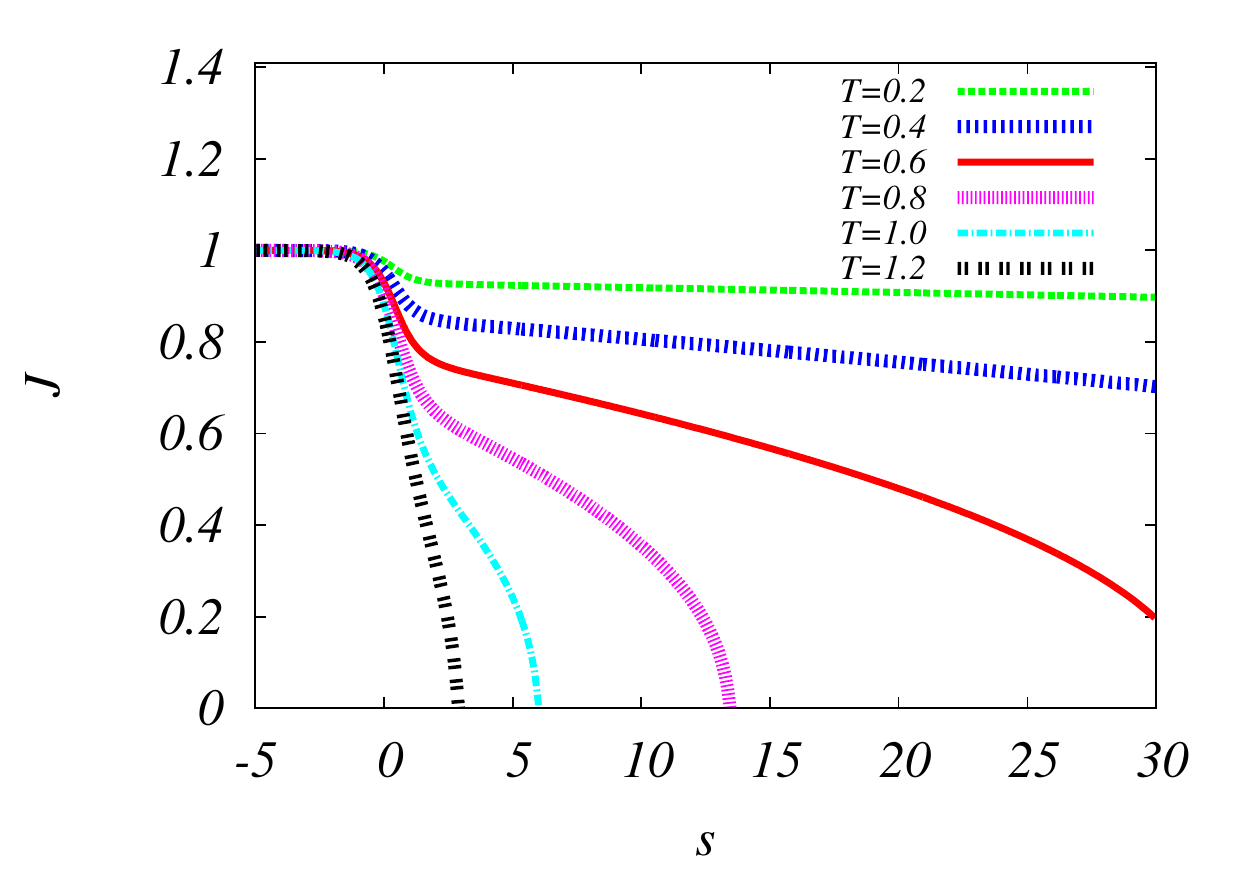}
\caption{Renormalized stiffness $J$ as a function of the logarithmic scale parameter $s$ for several choices of the temperature.}
\end{center}
\end{figure}

\subsubsection{Numerical solution}

We now complement the analytic results by an explicit numerical solution of the flow equations for a specific choice of the initial parameters in the bare action at various temperatures. The model parameters are $\alf_0 = 1$, $Z_0 = 1$, and $u_0 = 1$. The ultraviolet momentum cutoff is $\Lam_{\rm uv} = 1$.
For the regulator we choose a smooth exponential momentum cutoff \cite{jakubczyk14}
\begin{equation} \label{regulator}
 R(\bq) = Z_\pi \bq^2 r(\bq^2/\Lam^2) \;\; \mbox{with} \;\;
 r(x) = \frac{2}{e^x - 1} \, .
\end{equation}
We start the flow at $\Lam = \Lam_0 = e^5 \gg \Lam_{\rm uv}$, such that initially all fluctuations are practically suppressed.
Flows will be presented as a function of the logarithmic flow parameter
$s = \ln{\Lam_{\rm uv}/\Lam}$.
For $\Lam > \Lam_{\rm uv}$, corresponding to $s<0$, there are only small contributions to the flow.

In Fig.~1 we show the flow of the stiffness $J$ for various temperatures.
At low temperatures one can clearly see a large regime where the stiffness decreases logarithmically as a function of $\Lam$, that is, linearly as a function of $s$, in agreement with Eq.~(\ref{J_log}). The slope is proportional to $T^2$ as predicted (see again Eq.~(\ref{J_log})). For larger temperatures the regime controlled by the quasi-fixed point shrinks and the complete collapse of $J$ is shifted to larger scales $\Lam$, that is, smaller $s$.

\begin{figure}[tb]
\begin{center}
\includegraphics[width=7cm]{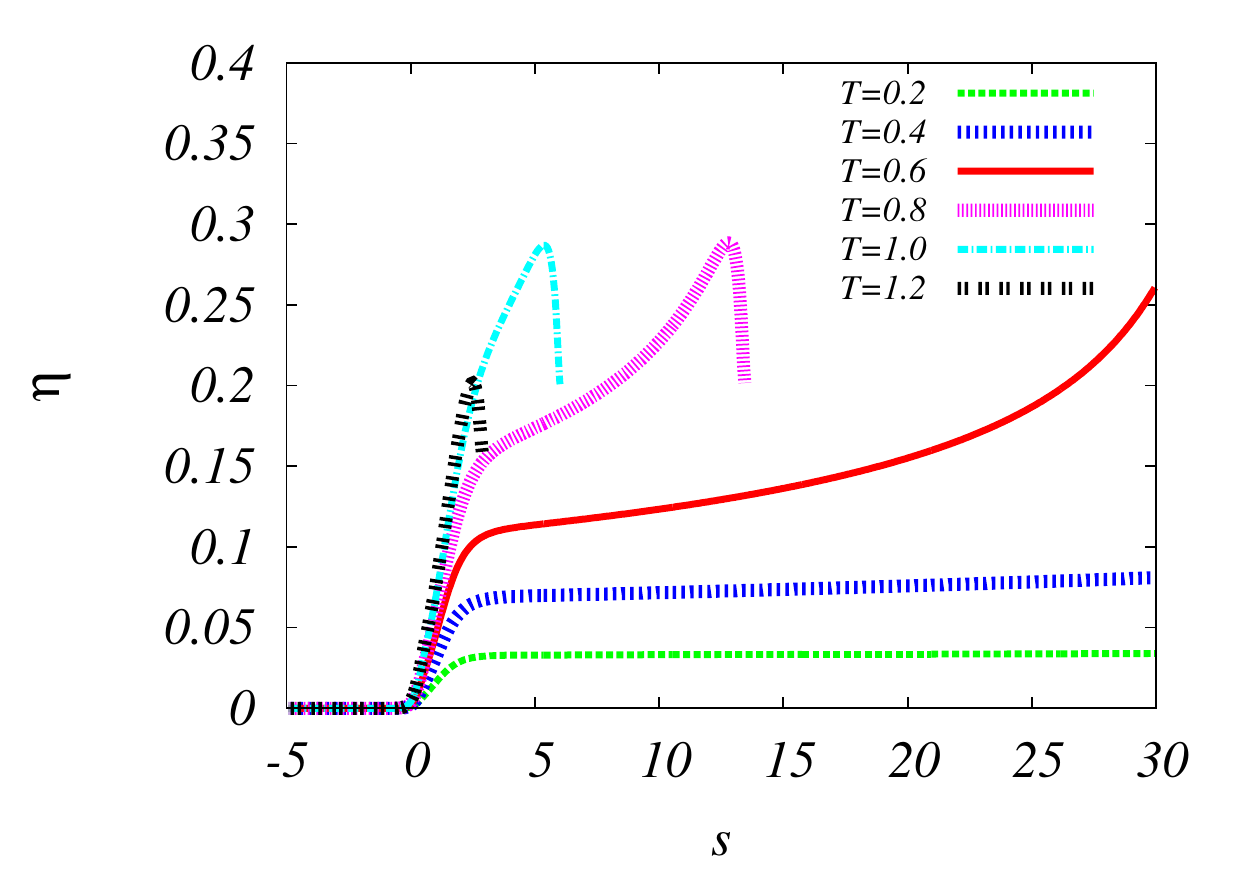} \\
\includegraphics[width=7cm]{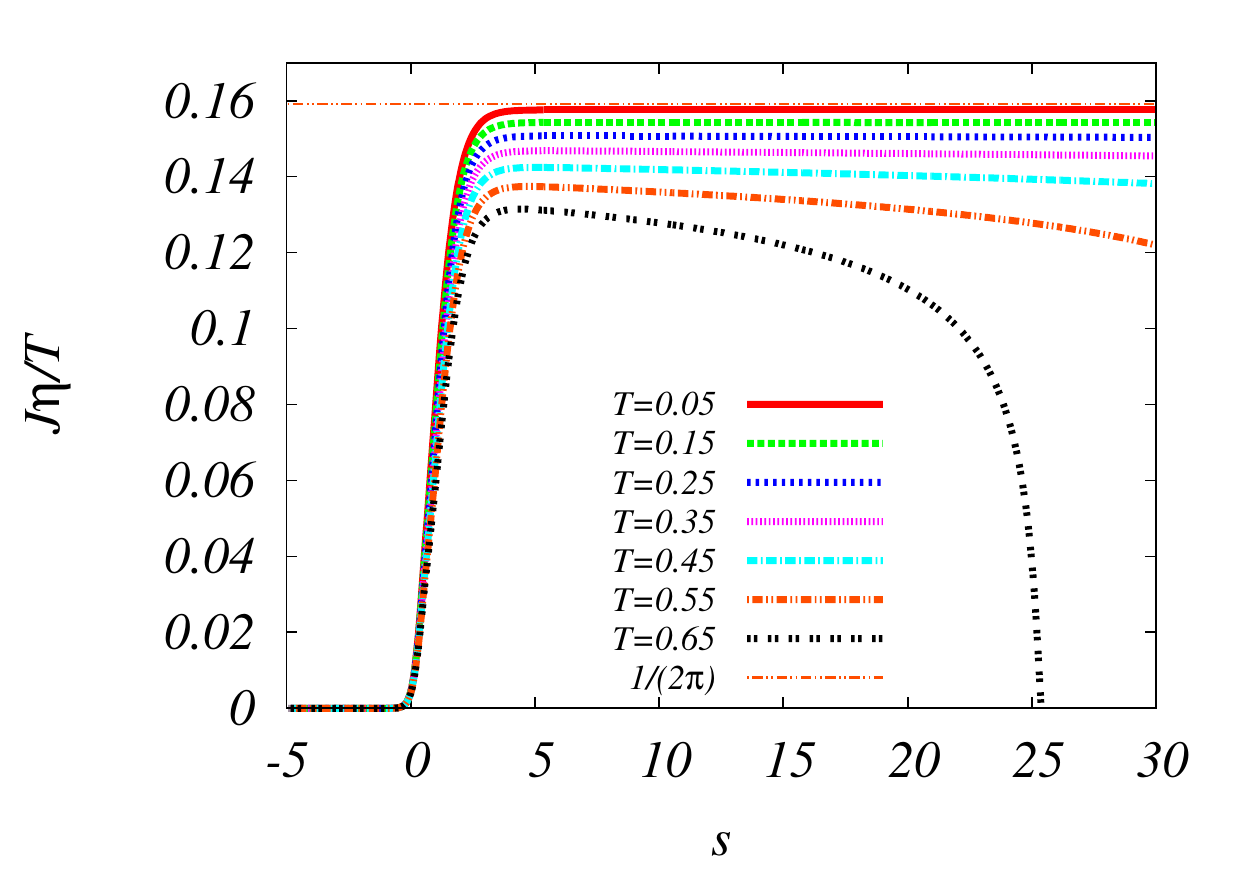}
\caption{Top: Anomalous dimension $\eta$ as a function of the scale parameter $s$ for various choices of the temperature. Bottom: $J\eta/T$ as a function of $s$ for various temperatures. The horizontal line represents the low temperature limit $1/(2\pi)$.}
\end{center}
\end{figure}
The flow of the anomalous dimension $\eta$ is shown in Fig.~2. At low temperatures it quickly reaches a quasi-plateau where it increases only logarithmically as a function of decreasing $\Lam$. In that regime $\eta$ is inversely proportional to $J$, that is, the product $\eta J$ is invariant. The ratio $J\eta/T$ is temperature dependent, in agreement with Eq.~(\ref{eta_mod}), since the numerical results have been obtained with a regulator function proportional to $Z_\pi$, see Eq.~(\ref{regulator}).

\begin{figure}[tb]
\begin{center}
\includegraphics[width=7cm]{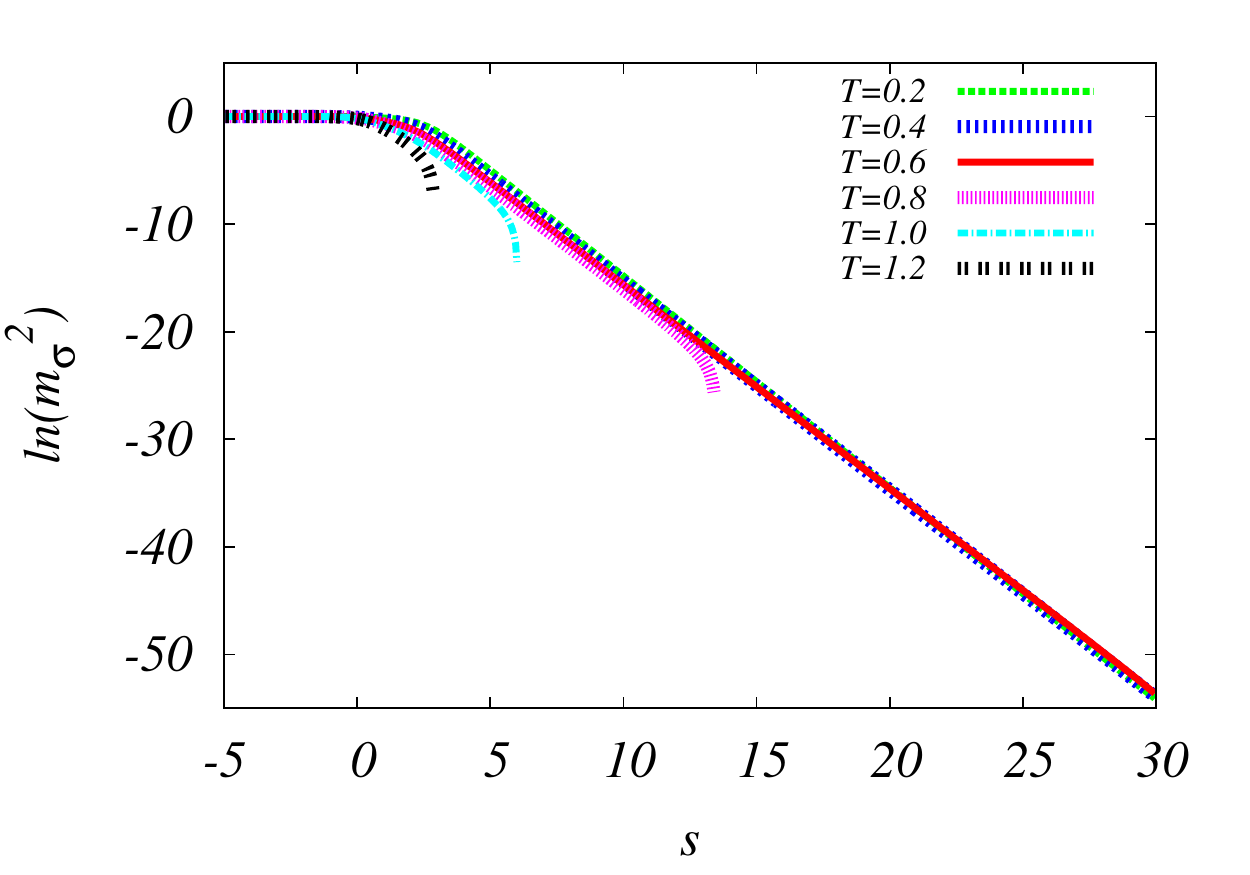}
\caption{Longitudinal mass $m_\sg^2$ as a function of the scale parameter $s$ for various choices of the temperature.}
\end{center}
\end{figure}
The rapid decrease of the longitudinal mass $m_\sg^2$ as a function of decreasing $\Lam$ is evident from Fig.~3. For temperatures below $T=1.0$ there is an extended regime where the flow follows the power law decay $m_\sg^2 \propto \Lam^{2-\eta}$, in agreement with Eq.~(\ref{m_sg}). One can also see that the straight line in the power law regime is shifted upwards at lower temperatures, in agreement with the factor $T^{-1}$ in Eq.~(\ref{m_sg}).

\begin{figure}[b]
\begin{center}
\includegraphics[width=7cm]{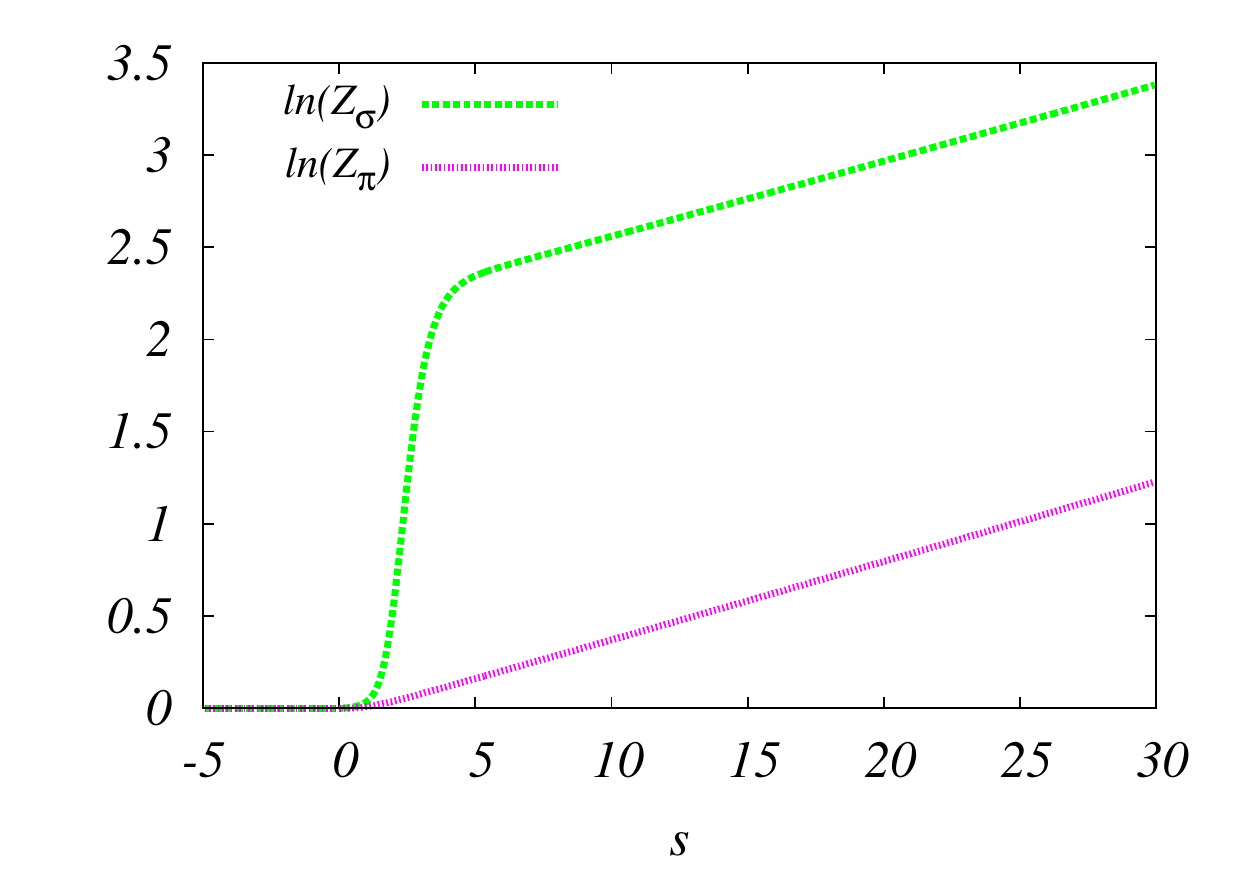} \\
\includegraphics[width=7cm]{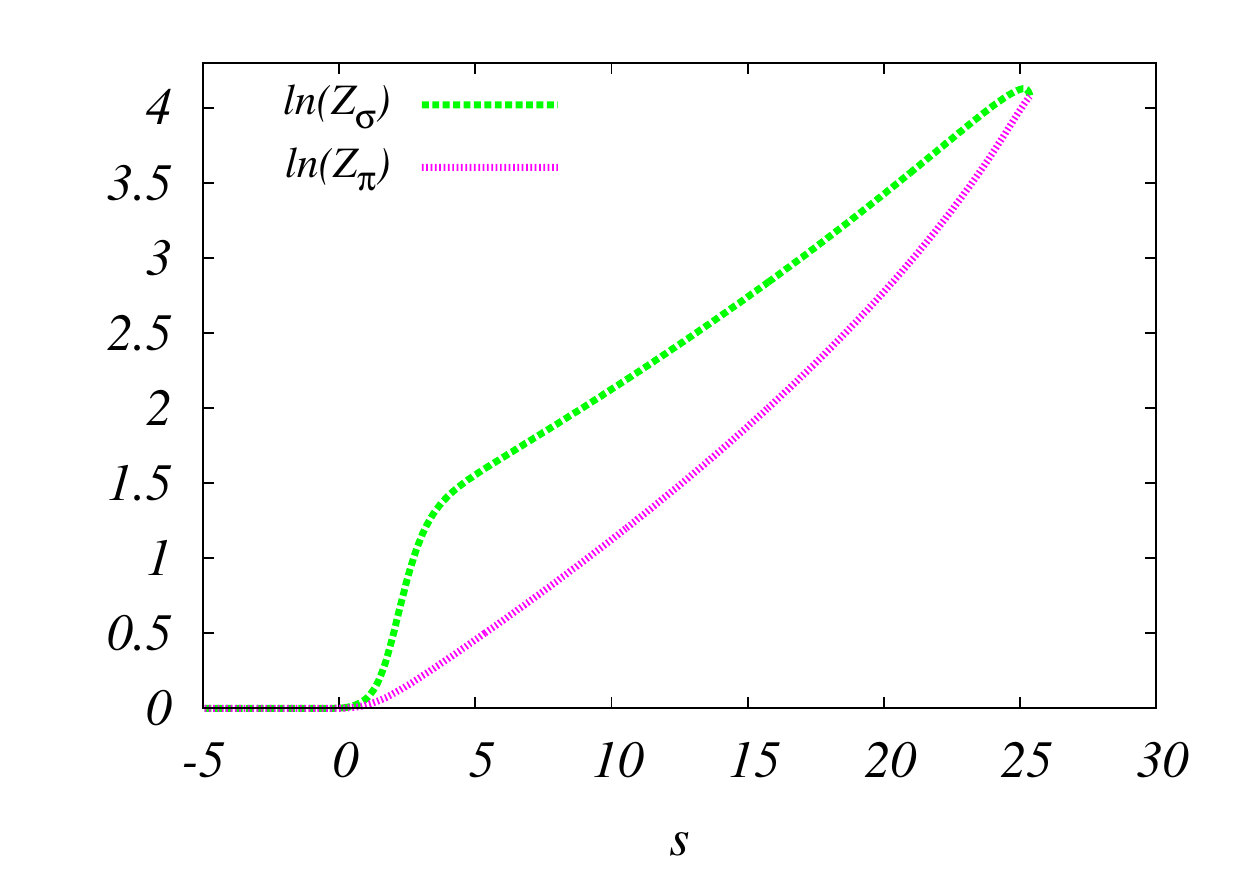}
\caption{Double-logarithmic plot of the $Z$-factors $Z_\sg$ and $Z_\pi$ as a function of the scale for two distinct temperatures $T=0.25$ (top) and $T=0.65$ (bottom).}
\end{center}
\end{figure}
The flow of the $Z$-factors for longitudinal and transverse fluctuations is shown for two distinct temperatures in Fig.~4. Note that $Z_\sg$ is always larger than $Z_\pi$, until the point where the stiffness $J$ collapses. At the lower temperature $T = 0.25$ one can see the power laws $Z_\pi, Z_\sg \propto \Lam^{-\eta}$ in agreement with Eqs.~(\ref{Z_pi}) and (\ref{Z_sg}). The ratio $Z_\sg/Z_\pi$ is proportional to $T^{-1}$ in the quasi-fixed point regime at low temperatures, as expected from Eq.~(\ref{Z_sg}). It thus diverges for $T \to 0$. This matches with the behavior of the $Z$-factors in the ground state of the interacting Bose gas, where $Z_\sg$ diverges due to transverse quantum fluctuations, while $Z_\pi$ remains finite.\cite{obert13}

\begin{figure}[tb]
\begin{center}
\includegraphics[width=7cm]{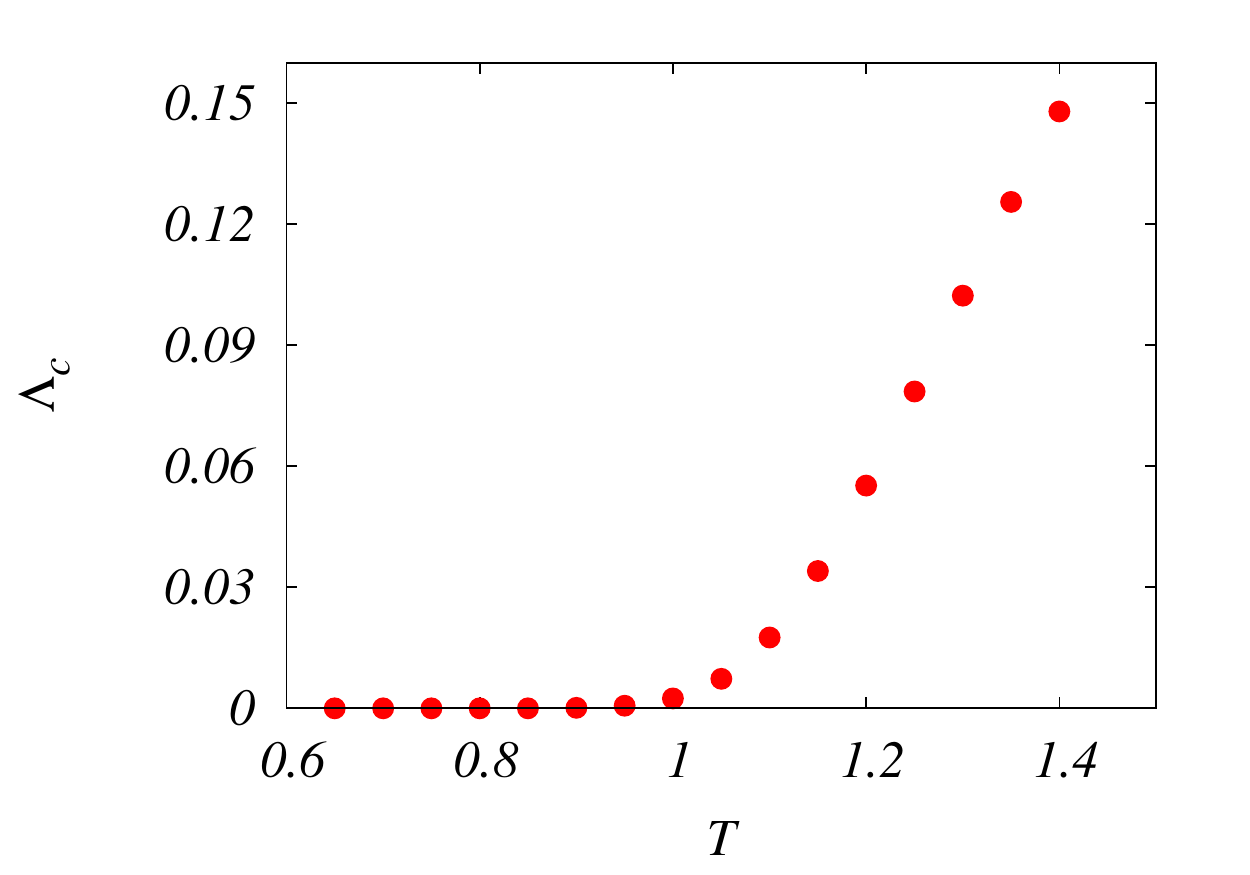}
\caption{Critical scale $\Lam_c$ as a function of temperature.}
\end{center}
\end{figure}
The scale $\Lam_c$ at which $J$ vanishes is shown as a function of temperature in Fig.~5. The exponential decrease upon lowering $T$ is clearly visible. For $T < 1$ the critical scale is tiny. The corresponding correlation length $\xi \sim \Lam_c^{-1}$ is huge and becomes quickly practically infinite for $T \ll 1$.

%%%%%%%%%%%%%%%%%%%%%%%%%%%%%%%%%%%%%%%%%%%%%%%%%%%%%%%%%%%%%%%%%%%%%%%%%%%%%%

\section{Density-phase representation}

In the preceding sections, the flow was derived for a linear (cartesian) decomposition of the complex field $\phi$ in longitudinal and transverse components, see Eq.~(\ref{cartesian}). Alternatively, one may decompose the field in {\em polar coordinates}, that is, amplitude and phase, and express the amplitude as a square root of the superfluid density, \cite{popov87}
\begin{equation}
 \phi(\br) = \sqrt{\rho(\br)} e^{i\vartheta(\br)} \, .
\end{equation}
Inserting this parametrization into the bare action (\ref{cS}), and splitting $\rho(\br) = \rho_0 + \tilde\rho(\br)$ with $\rho_0 = \alf_0^2$, one obtains
\begin{eqnarray} \label{cS_polar}
 \cS[\phi] &=& \beta \int d^2\br \, \Big[ \,
 \frac{u_0}{8} \tilde\rho^2(\br) +
 \frac{Z_0 \rho_0}{2} \big(\nabla\vartheta(\br)\big)^2 \nonumber \\
 &+& \frac{Z_0}{2} \tilde\rho(\br) \big(\nabla\vartheta(\br)\big)^2 +
 \frac{Z_0}{8} \frac{\big(\nabla\tilde\rho(\br)\big)^2}{\rho_0 + \tilde\rho(\br)}
 \, \Big] \, ,
\end{eqnarray}
The Jacobian associated with the transformation from $\Re\phi$ and $\Im\phi$ to the variables $\rho$ and $\vartheta$ is one, and thus does not lead to additional terms in the functional integral.
Note that the quartic interaction term in the action (\ref{cS}) has been transformed to a quadratic density fluctuation term, while the quadratic gradient term in (\ref{cS}) gives rise to a quadratic phase-gradient term, but also to interaction terms.

A distinctive advantage of the density-phase representation is that all interaction terms involve gradients, such that the interaction vertices are strongly suppressed at small momenta. Hence, the perturbation expansion is free from infrared divergences.\cite{popov87}
The first term in Eq.~(\ref{cS_polar}) provides the density fluctuations with a mass. Due to the absence of infrared divergences this mass is not affected significantly by phase fluctuations. This is very different from the fate of the longitudinal mass discussed in the preceding sections.

Let us briefly review how the algebraic decay of order parameter correlations is obtained from the functional integral in density-phase coordinates,\cite{popov87}
by evaluating the correlation function
\begin{equation}
 G(\br) = \bra \phi(\br) \phi^*(\b0) \ket =
 \bra \sqrt{\rho(\br) \rho(\b0)} e^{i[\vartheta(\br) - \vartheta(\b0)]} \ket
\end{equation}
at long distances $r = |\br|$.
The density fluctuations are not expected to have any qualitative effect at long distances, since they are massive. Setting $\tilde\rho(\br) = 0$ one can replace $\rho(\br)$ by $\rho_0$ and the action is reduced to the quadratic phase fluctuation term. This yields \cite{popov87}
\begin{equation}
 G(\br) =
 \rho_0 e^{-\frac{1}{2} \bra [ \vartheta(\br) - \vartheta(\b0) ]^2 \ket} \, .
\end{equation}
Using the Fourier representation
$\vartheta(\br) = T \int_\bq \vartheta_\bq e^{i\bq\br}$, one can express the expectation value in the exponent as
\begin{equation}
 \bra [ \vartheta(\br) - \vartheta(\b0) ]^2 \ket =
 T \int_\bq G_\vartheta(\bq) |e^{i\bq\br} - 1|^2 \, ,
\end{equation}
where
\begin{equation}
 G_\vartheta(\bq) = \frac{1}{J \bq^2} \, ,
\end{equation}
with the phase stiffness $J = Z_0 \rho_0$.
Representing $\bq$ by polar coordinates $q$ and $\varphi$, one can write $\bq\br = qr \cos\varphi$. Performing the integration over the angle $\varphi$, one obtains
\begin{equation} \label{Gr}
 G(\br) =
 \rho_0 \exp \left[ - \eta \int_0^{\Lam_{\rm uv}} \! dq \, \frac{1 - J_0(qr)}{q}
 \right]
 \, ,
\end{equation}
where $\Lam_{\rm uv}$ is the ultraviolet cutoff, and
\begin{equation} \label{eta1}
 \eta = \frac{T}{2\pi J} \, .
\end{equation}
The Bessel function is defined as
$J_0(x) = \int_0^{2\pi} \frac{d\varphi}{2\pi} \cos(x \cos\varphi)$.
The integral is convergent at small $q$ since $J_0(0) = 1$.
At large $q$ the integral diverges logarithmically,
$\int_0^{\Lam_{\rm uv}} \! dq \, \frac{1 - J_0(qr)}{q} =
 \int_0^{\Lam_{\rm uv} r} \! dx \, \frac{1- J_0(x)}{x}$ 
$\stackrel{r \to \infty}{\to} \ln(\Lam_{\rm uv} r) + \mbox{const}$,
so that
\begin{equation}
 G(\br) \propto \rho_0/r^\eta \quad \mbox{for} \quad
 r \to \infty \, .
\end{equation}
This is the celebrated algebraic decay of order parameter correlations in the BKT phase.

Since there are no infrared divergences in the density-phase representation, there is no need to introduce infrared regulators and to compute a renormalization group flow.
Nevertheless, to make a connection with the flow computed for the cartesian decomposition of the order parameter field, we now extract the flow of $\alf$, $m_\sg^2$, etc.\ from the density-phase representation in the presence of an infrared cutoff $\Lam$ acting on the phase variable $\vartheta$.

With a sharp infrared cutoff on $\vartheta_\bq$, equation (\ref{Gr}) for $G(\br)$ is modified to
\begin{equation} \label{GLamr}
 G^\Lam(\br) =
 \rho_0 \exp \left[ - \eta \int_\Lam^{\Lam_{\rm uv}} \! dq \, \frac{1 - J_0(qr)}{q}
 \right] \, .
\end{equation}
For $\Lam > 0$, there is also an {\em anomalous}\/ correlation function
\begin{equation}
 F^\Lam(\br) = \bra \phi(\br) \phi(\b0) \ket =
 \bra \sqrt{\rho(\br) \rho(\b0)} e^{i[\vartheta(\br) + \vartheta(\b0)]} \ket \, .
\end{equation}
Following the steps leading to Eq.~(\ref{GLamr}) for $G^\Lam(\br)$, and choosing the average phase $\bra \vartheta(\br) \ket$ equal to zero, one obtains
\begin{equation} \label{FLamr}
 F^\Lam(\br) =
 \rho_0 \exp \left[ - \eta \int_\Lam^{\Lam_{\rm uv}} \! dq \, \frac{1 + J_0(qr)}{q}
 \right] \, .
\end{equation}
Note that $F^\Lam(\br)$ vanishes for $\Lam \to 0$, for any $\br$, since the integral in the exponent is logarithmically infrared divergent.
Using the cartesian decomposition (\ref{cartesian}) of $\phi(\br)$, we can relate $G^\Lam(\br)$ and $F^\Lam(\br)$ to $G_\sg^\Lam(\br)$, $G_\pi^\Lam(\br)$ and $\alf^\Lam$ by the linear expressions
\begin{eqnarray} \label{GFGsgGpi}
 G^\Lam(\br) &=& (\alf^\Lam)^2 + G_\sg^\Lam(\br) + G_\pi^\Lam(\br) \, , \\
 F^\Lam(\br) &=& (\alf^\Lam)^2 + G_\sg^\Lam(\br) - G_\pi^\Lam(\br) \, .
\end{eqnarray}

The order parameter $\alf^\Lam$ can be extracted from the long-distance limit of $G^\Lam(\br)$ or $F^\Lam(\br)$. Since $J_0(qr)$ vanishes for $r \to \infty$, one obtains
\begin{equation} \label{rhoLam}
 \rho^\Lam = (\alf^\Lam)^2 = \rho_0 (\Lam/\Lam_{\rm uv})^\eta \, .
\end{equation}
This agrees with the result (\ref{alf}) obtained for $\alf^2$ in Sec.~IV. The slight difference in the prefactor is due to the different form of the momentum cutoff chosen here and there.

Using Eq.~(\ref{rhoLam}), one can write the expressions (\ref{GLamr}) and (\ref{FLamr}) as
\begin{eqnarray}
 G^\Lam(\br) &=& \rho^\Lam \exp \left[ \phantom - \eta \int_\Lam^{\Lam_{\rm uv}} \! dq
 \frac{J_0(qr)}{q} \right] =
 \rho^\Lam e^{G_\vartheta^\Lam(\br)} \, , \\
 F^\Lam(\br) &=& \rho^\Lam \exp \left[ - \eta \int_\Lam^{\Lam_{\rm uv}} \! dq
 \frac{J_0(qr)}{q} \right] =
 \rho^\Lam e^{-G_\vartheta^\Lam(\br)} \, ,
\end{eqnarray}
where
\begin{equation}
 G_\vartheta^\Lam(\br) = T \int_{\Lam \leq |\bq| \leq \Lam_{\rm uv}}
 \frac{d^2\bq}{(2\pi)^2} \frac{1}{J \bq^2} e^{i\bq\br} \, .
\end{equation}
Inverting (\ref{GFGsgGpi}) thus yields
\begin{eqnarray}
\label{G_piG_th}
 G_\pi^\Lam(\br) &=& \rho^\Lam \sinh G_\vartheta^\Lam(\br) \, , \\
\label{G_sgG_th}
 G_\sg^\Lam(\br) &=& \rho^\Lam \left[\cosh G_\vartheta^\Lam(\br) - 1 \right] \, .
\end{eqnarray}

We now compare the behavior of $G_\pi^\Lam(\bq)$ and $G_\sg^\Lam(\bq)$ at small momenta $\bq$ with the results obtained from the flow equations in Sec.~IV.
The dominant contributions at small $\bq$ come from contributions at large distances in real space. Since $G_\vartheta^\Lam(\br)$ decays at large $\br$, we can expand the hyperbolic functions to obtain, to leading order,
\begin{eqnarray}
\label{G_pi_exp}
 G_\pi^\Lam(\br) &\stackrel{r \to \infty}{\to}&
 \rho^\Lam G_\vartheta^\Lam(\br) \, , \\
\label{G_sg_exp}
 G_\sg^\Lam(\br) &\stackrel{r \to \infty}{\to}&
 \frac{1}{2} \rho^\Lam \big[ G_\vartheta^\Lam(\br) \big]^2 \, .
\end{eqnarray}
The asymptotic behavior of $G_\pi^\Lam(\br)$ is consistent with our ansatz (\ref{G_pi}) for $G_\pi^\Lam(\bq)$, if $Z_\pi^\Lam \propto \Lam^{-\eta}$, in agreement with Eq.~(\ref{Z_pi}).
Fourier transforming the relation (\ref{G_sg_exp}) between $G_\sg^\Lam(\br)$ and $G_\vartheta^\Lam(\br)$ in the large distance limit yields
\begin{equation} \label{thetabubble}
 G_\sg^\Lam(\bq) = \frac{T}{2} \, \rho^\Lam \!
 \int_\bp G_\vartheta^\Lam(\bp) G_\vartheta^\Lam(\bp+\bq) \, .
\end{equation}
The inverse longitudinal mass is thus obtained as
\begin{equation}
 (m_\sg^\Lam)^{-2} =  G_\sg^\Lam(\b0) = \frac{T}{2} \, \rho^\Lam \!
 \int_\bp G_\vartheta^\Lam(\bp) G_\vartheta^\Lam(\bp) \, .
\end{equation}
Note the similarity to the flow equation (\ref{floweq2_m_sg}).
Since the momentum integral diverges as $\Lam^{-2}$, and $\rho^\Lam \propto \Lam^\eta$, one obtains $(m_\sg^\Lam)^2 \propto T^{-1} \Lam^{2-\eta}$ in agreement with Eq.~(\ref{m_sg}).
Finally, comparing Eq.~(\ref{thetabubble}) with our ansatz (\ref{G_sg}) for $G_\sg^\Lam(\bq)$, one obtains $Z_\sg^\Lam \propto T^{-1} \Lam^{-\eta}$, in agreement with Eq.~(\ref{Z_sg}).

Hence, the power laws for $\alf^\Lam$, $Z_\pi^\Lam$, $m_\sg^\Lam$, and $Z_\sg^\Lam$ obtained from the flow derived in the cartesian decomposition of the order parameter field in Sec.~IV are all confirmed by the calculation in density-phase representation.
A discrepancy occurs if contributions from longitudinal fluctuations to the flow are taken into account, as these lead to a slow decrease of the stiffness $J$. Since it seems clear that density fluctuations can renormalize $J$ only by a finite amount, the flow of the stiffness found in the fRG calculations is most likely an artifact of the truncation.

%%%%%%%%%%%%%%%%%%%%%%%%%%%%%%%%%%%%%%%%%%%%%%%%%%%%%%%%%%%%%%%%%%%%%%%%%%%%%%

\section{Conclusion}

Motivated by previous fRG calculations, \cite{grater95,gersdorff01,jakubczyk14} we have investigated the interplay of longitudinal and transverse fluctuations in a $U(1)$-symmetric $\phi^4$-theory in two dimensions. A major goal was to clarify the mechanism leading to the unexpected decrease of the phase stiffness observed in the fRG flows.
To gain analytic insight, we used approximate flow equations obtained from an effective action that was truncated at quartic order in the fields and at quadratic order in gradients.
Discarding the longitudinal fluctuations we recovered the expected BKT fixed point with all its well-known properties: a finite phase stiffness, a vanishing order parameter, and an algebraic decay of order parameter correlations with an exponent $\eta$ proportional to $T$ at low temperatures.
Renormalized by transverse fluctuations, the longitudinal mass vanishes, so that longitudinal fluctuations become increasingly important for small momenta. Unlike the transverse fluctuations, their contribution to the stiffness does not vanish. In our truncation, the stiffness decreases logarithmically as a function of the momentum cutoff $\Lam$, leading to a collapse of the quasi-ordered BKT phase at a critical momentum scale $\Lam_c$, which corresponds to a finite correlation length $\xi \sim \Lam_c^{-1}$. The prefactor of the logarithmic decrease of $J$ is proportional to $T^2$. Hence, the final correlation length increases exponentially upon lowering the temperature.

The above results are consistent with more involved fRG calculations based on a non-perturbative derivative expansion of the effective action, \cite{gersdorff01,jakubczyk14} where the same slow decrease of the stiffness as in our calculation was found. Note, however, that these numerical flows suffer from singularities at $T \ll T_{\rm BKT}$, so that the asymptotic low temperature behavior cannot be extracted.

The decrease of the stiffness and the ensuing finite correlation length at any finite temperature in the fRG flows has always been interpreted as an artifact of the truncation. \cite{grater95,gersdorff01,jakubczyk14}
Indeed the algebraic decay of correlations in the BKT phase is very well established and confirmed by rigorous, numerical and experimental results.
In particular, in a famous work by Fr\"ohlich and Spencer \cite{froehlich81} the existence of a BKT-phase with a power-law decay of order parameter correlations at low temperatures was rigorously proven. Note, however, that the proof is restricted to systems having only a phase degree of freedom, such as the plane rotator or the Villain model.
To the best of our knowledge, there are no rigorous results on the BKT phase for two-dimensional systems with a complex order parameter consisting of phase and amplitude degrees of freedom such as the interacting Bose gas. \cite{cenatiempo14}

The flow of the stiffness obtained from the fRG is inconsistent with the results obtained from a density-phase representation of the order parameter field.\cite{popov87} To trace the origin of this discrepancy, we computed the flow of the parameters characterizing longitudinal and transverse fluctuations from a density-phase representation with a cutoff on phase fluctuations. Thereby, all power laws obtained from our truncated flow equations were confirmed. In particular, the longitudinal mass indeed scales to zero, with an exponent $2-\eta$. However, there is no flow of the stiffness in the density-phase representation. The logarithmic flow of the stiffness obtained in the cartesian decomposition of the order parameter field must therefore be canceled by higher order terms not included in our truncation.
Indeed, we already showed that the leading contributions to the stiffness flow (of order $T$) found in the work by Gr\"ater and Wetterich \cite{grater95} are canceled by including a density gradient term, so that the flow in our truncation arises only at order $T^2$.
A comparison to the low temperature expansion of the non-linear sigma model \cite{amit84} shows that the vanishing beta-function for the coupling in this model is thus reproduced to one-loop order in the truncation used by Gr\"ater and Wetterich, and to two-loop order in our truncation. The cancellation of terms of order $T^2$ requires a complete account of all
three-loop contributions in the non-linear sigma model. An early study of the scalar $\phi^4$-theory revealed that the usual field-theoretic loop expansion cannot be captured by a gradient expansion of the fRG effective action. \cite{papenbrock95} The latter corresponds to a Taylor expansion of momentum dependences.
Hence, we conclude that the unphysical flow of the stiffness found in the fRG calculations, including non-perturbative truncations based on the derivative expansion, is due to an insufficient parametrization of the momentum dependences of the effective action, which is rather hard to overcome.
Since the fRG is nevertheless a powerful approach, in particular for computing non-universal properties, one may adopt a pragmatic attitude and discard the small contributions leading to the logarithmic flow of the stiffness by hand.

%%%%%%%%%%%%%%%%%%%%%%%%%%%%%%%%%%%%%%%%%%%%%%%%%%%%%%%%%%%%%%%%%%%%%%%

\begin{acknowledgments}
We are grateful to A.~Alastuey, A.~Auerbach, K.~Byczuk, C.~Castellani, C.~Di Castro, B.~Delamotte, N.~Dupuis, A.~Eberlein, T.~Enss, T.~Holder, H.~Kn\"orrer, J.~Pawlowski, J.~Piasecki, H.~Spohn, D.~Ueltschi, C.~Wetterich, R.~Zeyher, and especially to John Toner for valuable discussions.
PJ acknowledges support from the Polish National Science Center via grant
2014/15/B/ST3/02212.
\end{acknowledgments}

%%%%%%%%%%%%%%%%%%%%%%%%%%%%%%%%%%%%%%%%%%%%%%%%%%%%%%%%%%%%%%%%%%%%%%%

\end{document}